\documentclass[10pt,leqno]{article}
\usepackage{graphicx}
\usepackage{xcolor}
\usepackage{authblk}
\usepackage{caption}
\usepackage{subcaption}

\usepackage{indentfirst,csquotes}

\usepackage{amssymb,amsthm,amsmath}
\usepackage{xcolor,paralist,hyperref,titlesec,fancyhdr,etoolbox}

\hypersetup{ colorlinks=true, linkcolor=black, filecolor=black, urlcolor=black }

\usepackage{lipsum}

\newcommand{\editnote}[2]{}
\title{Physics Prospects for a near-term Proton-Proton Collider} %%%%%%%%%%%%
\author[1]{Viviana Cavaliere}
\author[2]{Monica Dunford}
\author[3,4]{Heather M. Gray}
\author[5]{Elliot Lipeles}
\author[6]{Alison Lister}
\author[7,8]{Clara Nellist}

\affil[1]{Brookhaven National Laboratory}
\affil[2]{Kirchhoff-Institut f\"{u}r Physik, Universität Heidelberg}
\affil[3]{Lawrence Berkeley National Laboratory}
\affil[4]{University of California, Berkeley}
\affil[5]{University of Pennsylvania}
\affil[6]{University of British Columbia}
\affil[7]{University of Amsterdam}
\affil[8]{Nikhef}

\begin{document}

\date{\today}
\maketitle
 % \begin{figure}
 %     \centering
 %     \includegraphics[width=1.2\textwidth]{figs/Endorsements.pdf}
 %     \caption{Endorsments as of 30/03}
 %     \label{fig:enter-label}
 % \end{figure}

%\clearpage
\begin{abstract}
Hadron colliders at the energy frontier offer significant discovery potential through precise measurements of Standard Model processes and direct searches for new particles and interactions. A future hadron collider would enhance the exploration of particle physics at the electroweak scale and beyond, potentially uniting the community around a common project. The LHC has already demonstrated precision measurement and new physics search capabilities well beyond its original design goals and the HL-LHC will continue to usher in new advancements. 
This document highlights the physics potential of an FCC-hh machine to directly follow the HL-LHC. In order to reduce the timeline and costs, the physics impact of lower collider energies, down to $\sim 50$~TeV, is evaluated. Lower centre-of-mass energy could leverage advanced magnet technology to reduce both the cost and time to the next hadron collider.
Such a machine offers a breadth of physics potential and would make key advancements in Higgs measurements, direct particle production searches, and high-energy tests of Standard Model processes. Most projected results from such a hadron-hadron collider are superior to or competitive with other proposed accelerator projects and this option offers unparalleled physics breadth. The FCC program should lay out a decision-making process that evaluates in detail options for proceeding directly to a hadron collider, including the possibility of reducing energy targets and staging the magnet installation to spread out the cost profile.
\end{abstract} %%%%%%%%%

\section{Introduction}

This document proposes an FCC-hh machine to directly follow the HL-LHC. In order to reduce the timeline and costs, the physics impact of lower collider energies, ranging down to $\sim 50$~TeV, is evaluated.

Hadron colliders at the energy frontier provide tremendous discovery potential both through precision measurements of Standard Model processes and through searches for the direct production of new particles. 
A future hadron collider would provide the community with a rich and attractive program to explore particle physics at the electroweak scale and beyond and has the potential to unite the community behind a single goal. The precision measurement and new physics search potential have been demonstrated at hadron colliders with the SppS, Tevatron, and now the LHC. This will be further explored at the HL-LHC. At the LHC, improvements to analyses have repeatedly surpassed even optimistic predictions from the LHC community.

The physics capabilities of hadron colliders depend primarily on the energy and the luminosity, and the costs and timescales depend most strongly on the magnet field strength and the tunnel size. The original FCC-hh proposal centred on a 100~TeV machine in a 100\, km tunnel using 16~T dipoles with a luminosity sufficient to achieve 30\,ab$^{-1}$ over two experiments~\cite{FCC:2019vol3} as a second stage after the FCC-ee. The FCC-hh baseline has recently been updated to an 84~TeV machine using 14~T magnets~\cite{FCCUpdate}. A recent report on the status of the magnet development for the FCC gives 2055 as a possible start date for a 90~TeV FCC-hh collider, and that 5-10 years could be gained by anticipating some phases and through strong cooperation with industrial partners~\cite{MagnetTimeLine}. 
In Section~\ref{collider} we address the impacts of reducing the energy target ranging as low as $\sim 50$~TeV, which can be expected to make an early start date even more feasible both technically and financially.

%In this document, we introduce the idea of building an FCC-hh, with a slightly reduced, energy, directly after the HL-LHC. The energy of such a machine could range from 50 to 84~T, with the goal being to exploit existing magnet technology to reduce both the cost and the timeline. % AL removed as repeat of abstract

%Current estimates for a future hadron collider (FCC-hh) suggest that a target energy of $50-70$ TeV is technically achievable on a 2055 timescale. This is detailed in th  e FCC proposal. 

%It currently seems feasible to use LHC magnets within the FCC tunnel and reach a centre-of-mass energy of \~50~TeV on the timescale of 2045 (in an ideal funding scenario).

This document highlights selected key areas where hadron colliders can make powerful contributions to our understanding of particle physics. Section \ref{higgs} presents a summary of expected Higgs measurements in the context of a hadron machine fulfilling a Higgs factory role. 
Section \ref{direct_production} reviews the power of direct searches for new particles and Section \ref{precisionvproduction} compares them to indirect constraints. 
Hadron colliders provide a large breadth of experimental tests (e.g. tests of the behaviour of SM processes at very high energies), the reach of which can be studied via a coherent framework provided by effective field theory as discussed in Section \ref{eft}. 
%While comparison can be made between Higgs coupling and other electroweak measurements, 

\section{Broad Energy and Luminosity Considerations}
\label{collider}

The energy reach of hadron colliders is primarily determined by tunnel size and the magnet field strength, There are two major additional constraints in this energy regime. Firstly, the synchrotron radiation from the stored beam must be low enough to be extracted from the cryogenic environment of the beam pipe\cite{Zimmermann} and, secondly, the detectors must be capable of handling the associated pile-up. The corresponding limits are 2 kW/beam for the synchrotron radiation \cite{Zimmermann}, and currently assumed to be 1000 for the pile-up, although detailed detector studies in this regime still need to be performed.

Table \ref{tab:Zimmermann} adapted from Reference \cite{Zimmermann} shows three possible scenarios with high-field magnets: F12PU, F14 and F17, with each scenarios labelled by the dipole field strength. In addition, we show the parameters for the HL-LHC and an additional scenario with an energy of 50\,TeV, which would use magnets with the same field strength as current LHC magnets. All scenarios would use the new 91\, km FCC tunnel. Scenarios F12PU and 50 TeV are pile-up limited and consequently include luminosity levelling to not exceed $\approx 1000$ collisions per crossing. The two constraints cross around F14, which means that  F17 would be synchrotron radiation limited. 

%\editnote{MD}{Elliot/Heather: Comment from Andreas 'I do not understand the motivation for giving “Initial events / crossing” in Table instead of the nominal (or ultimate) numbers. I pasted below the latest (but still preliminary) HL-LHC luminosity scenario.'}
%HG: I agree with Andreas. We should be using the ultimate numbers to compare. Is 135 the real pile up number now instead of 140 or 200?

\begin{table}[]
    \centering
    \begin{tabular}{lcccccc}
    \hline\hline
                     &        & HL-LHC \cite{ZurbanoFernandez:2020cco}  &  &  &    &   \\
    Parameter                 & Unit      & initial (ultimate) & 50 TeV & F12PU & F14   & F17  \\
    \hline
    Centre-of-mass energy     & TeV       & 14 & 50  & 72    & 84    & 102    \\
    Peak arc dipole field     & T         & 8.3  & 8.3 & 12    & 14    & 17     \\
    SR power / beam           & W        & 7.3 & & 1450  & 1200  & 2670     \\
    Peak Collisions / crossing& -         & 135 (200) & 1000   & 1000   & 920 & 975    \\
    Luminosity / yr           & fb$^{-1}$ & 240 (350) & 1300 & 1300 & 920 & 920   \\
    \hline\hline
    \end{tabular}    
    \caption{Scenarios for Hadron Colliders in the FCC tunnel. F12PU, F14, and F17 are from Reference\cite{Zimmermann}. The 50 TeV has been added as an additional comparison point. HL-LHC numbers have been updated to ultimate luminosity numbers. }
    \label{tab:Zimmermann}
\end{table}

A detailed assessment on the timescale of such a collider needs to be performed. However as discussed in the introduction, a physics production by the 2050's would be feasible.

A successful outcome of this proposal would allow a significantly broader physics program to be performed on an earlier timescale. However, the costs of such a machine are significant and the financial feasibility of the project needs to be studied carefully. For example, it might be possible to reduce the cost profile by staging the magnet installation, e.g. by installing half the magnets initially to reach half the energy, even at the cost of an increase to the total project cost. A hadron collider directly following the LHC would reduce the total FCC project cost by removing the costs associated with FCC-ee.  

%\editnote{MD}{Elliot: comment from Andreas "I think there is overall agreement in the community that a high-energy hadron collider is much more powerful than an e+e- machine, so the question is only when to build it. You advocate a fast track with the compromise on the energy, and you make good points why 70 TeV may be sufficient. A problem is that this still requires Nb3Sn technology, which is terrible to handle (the FCC-hh studies assume those magnets can be built at the same cost as the current NbTi ones, which will be very hard to achieve). You will need to spend 20B or more, and you can spend this much money only once (and you need a very strong case to get it!). So, many people will argue that we should first complete our R&D before embarking into this, and not rush. I am not sure this point is sufficiently addressed in your document". Not sure if you want to try to address that here. }

\section{Higgs Physics}
\label{higgs}
Fully characterizing the physics of the Higgs boson is an important component in the physics program of any future collider. Some of the key benefits to hadron colliders are its large sample of Higgs bosons, which enable precise measurements of Higgs coupling, its ability to probe Higgs self-coupling, which is important for understanding the Higgs potential and its role in electroweak symmetry breaking, and to address questions related to Higgs mass stabilization and naturalness, which are critical to exploring extensions of the Standard Model such as top partners, supersymmetry (SUSY), and compositeness. 

\begin{figure}
    \centering
    \includegraphics[width=0.45\textwidth]{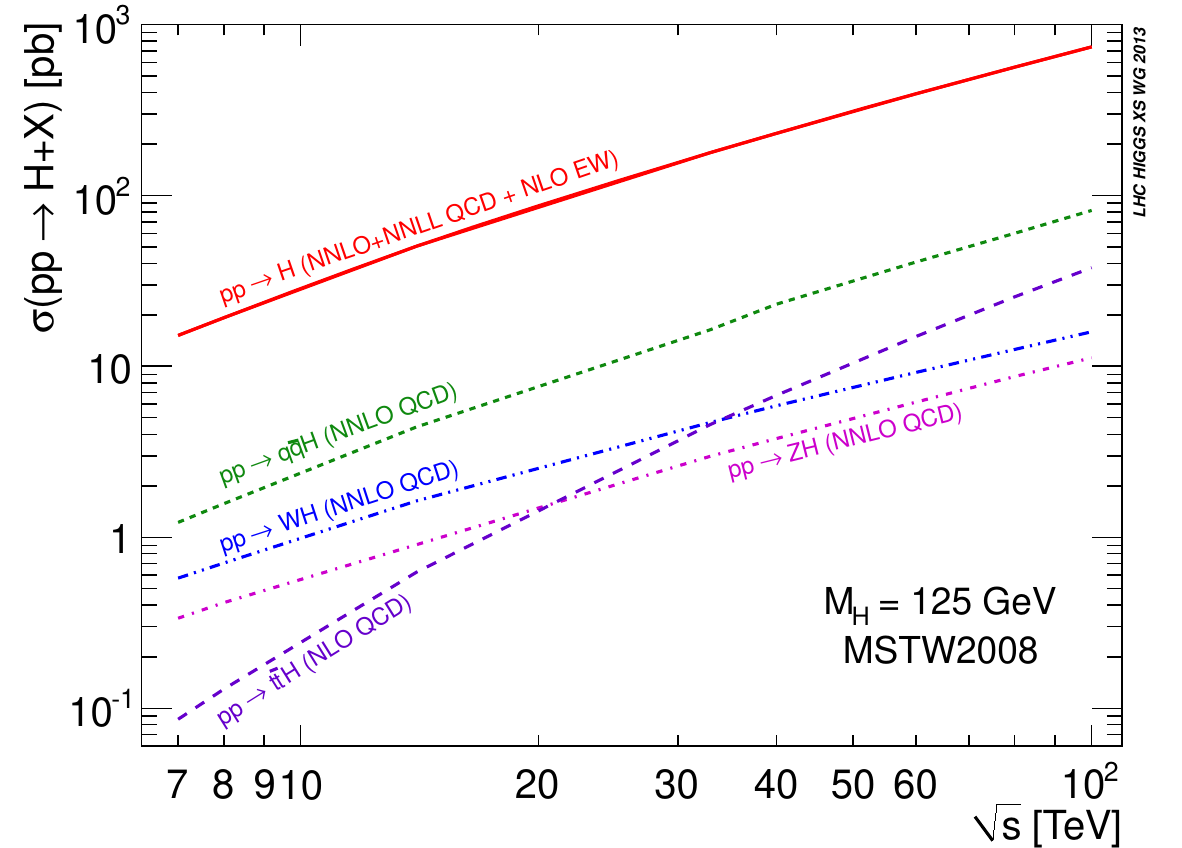}
    \includegraphics[width=0.5\textwidth]{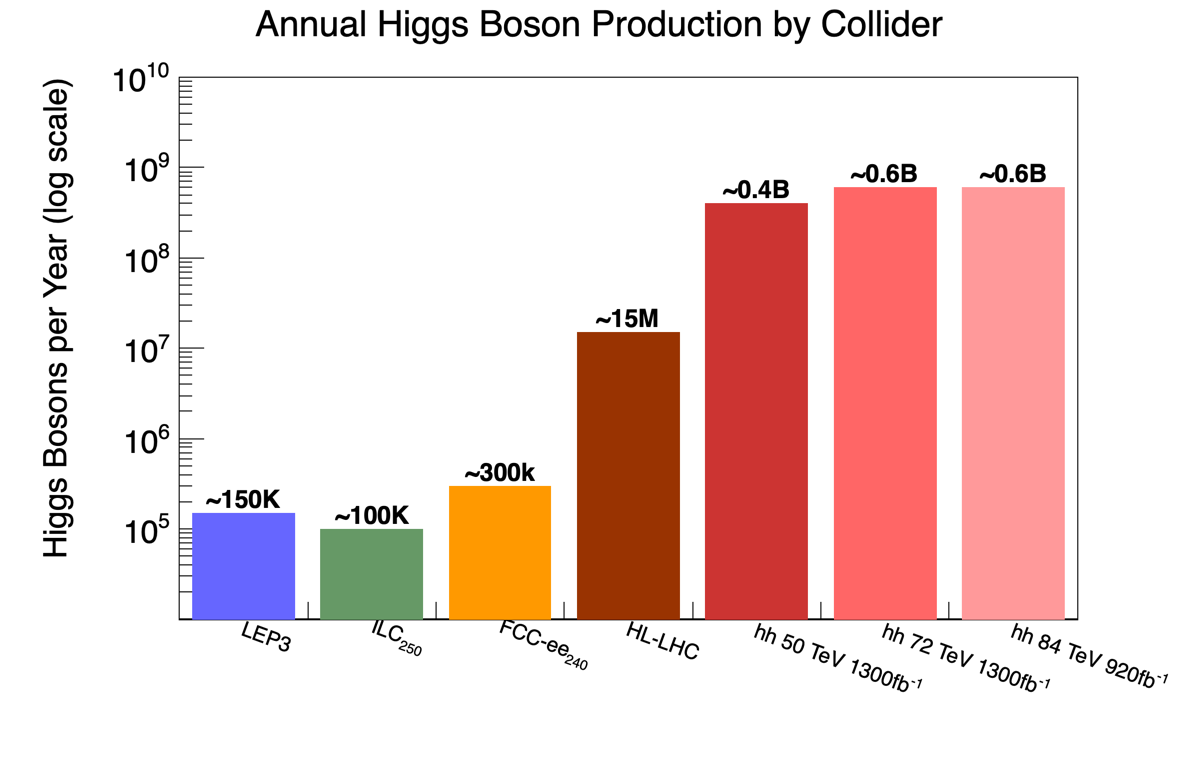}
    \caption{Left: Higgs-boson production cross sections as a function of centre-of-mass energies from Ref.~\cite{HiggsEuropeanStrategy}. Right: Annual production of Higgs bosons per collider option per experiment.}
    \label{fig:higgsxs}
\end{figure}

Figure~\ref{fig:higgsxs} shows the production cross-section for the Higgs boson as a function of the centre-of-mass energy of proton-proton colliders. All cross sections increase with energy but the increase is much larger up to 60~TeV and then more modest until 100~TeV due to the centre-of-mass energy of the system. % as shown in Table~\ref{tab:HiggsXsec}. 
Hadron colliders are the ultimate Higgs factory producing half a billion Higgs bosons per year for each experiment (Figure \ref{fig:higgsxs}). While only a fraction of these events will pass the selected analysis criteria, the statistical sample of Higgs bosons available for analysis at hadron colliders is on the order of hundreds of millions. The total number of Higgs bosons produced is similar for the three centre-of-mass energies shown here as the overall luminosity can compensate for a reduced centre-of-mass energy. 

\textit{}

%\begin{table}[h]
%    \centering
%    \begin{tabular}{lccc}
%    \hline\hline
%         Production Process & $\frac{\sigma(60 \mathrm{TeV})}{\sigma(14 \mathrm{TeV})}$ & $\frac{\sigma(100 \mathrm{TeV})}{\sigma(14 \mathrm{TeV})}$ \\
%        \hline
%        ggF  &  8x & 14x &  \\
%        ttH  &  24x & 40x &  \\
%        HH   &  18x &  42x &  \\
%\hline\hline
%    \end{tabular}
%    \caption{Expected increases in production cross-sections at higher collider energies.
%\editnote{HG}{Change this to 50, 70, 82 TeV.}}
%    \label{tab:HiggsXsec}
%\end{table}

\subsection{Higgs couplings and measurements}

 Accurately probing the Higgs coupling is essential for understanding the mechanism of electroweak symmetry breaking. While $e^+e^-$ colliders excel in model-independent Higgs coupling measurements, particularly through the determination of the Higgs width, hadron machines offer a broader range of precision measurements of Higgs coupling and are optimal for heavy final states and rare decays. These couplings can be measured across multiple channels and via measurements of Higgs coupling ratios can be model-independent probes, provided that theoretical uncertainties can be sufficiently understood. As seen in Table~\ref{tab:higgsKappas}, taken from Refs.~\cite{HLLHCUpdate} and~\cite{de_Blas_2020}, a hadron collider can provide sub-percent precision on all major couplings. Recent extrapolations show that a 70 (50)\, TeV collider could constrain $\kappa_b$ to 0.27 (0.28) and $\kappa_c$ to 2.2 (2.3) based on recent LHC analyses~\cite{ATLAS:2024yzu}. One of the significant advantages of a hadron machine is its ability to measure rare Higgs decays with 1000 times the cross-section compared to $e^+e^-$ colliders, making it a powerful tool for these rare processes, such as $Z\gamma$ decays as shown in the table. Higgs-to-invisible measurements can provide insights into the Higgs width.  The precision at both hadron colliders and $e^+e^-$ machines is predicted to be at the sub-percent level~\cite{FCC:2018byv}, with the former relying on a detailed understanding of the systematic uncertainties. While other exotic decays of the Higgs have not been studied, the very large Higgs production rate will likely lead to sub-percent sensitivities for the more challenging scenarios and even lower for the cleaner scenarios.
 
 %The precision at a hadron collider is predicted to be at the sub-percent level but not as competitive compared to $e^+ e^-$ machines, which is around 0.2\%.

%\editnote{MD}{Heather/Elliot: See Andreas's comments about Higgs to invisible. I have added the following line but I didn't find numbers for FCChh only. Do you hae them?} HG: rephrased to say 'sub-percent level for both and added a reference' and a comment about the systematics.

Due to the large statistical sample of Higgs boson produced, hadron colliders also provide precise measurements of differential distributions. In particular, the Higgs boson can be studied at high transverse momentum ($p_T$), which is a phase space that is sensitive to potential new physics. This capability allows for more precise measurements with reduced backgrounds. 
%Figure \editnote{MD}{Figure X demonstrates that the expected signal significance over the background reduces dramatically at high transverse momentum of the Higgs boson. 
Differential measurements provide an unprecedented kinematic reach and access to new physics at higher energy scales. Many Higgs measurements will be limited by theoretical uncertainties after the HL-LHC. Therefore, any successful physics programs relies on advancements in theory in addition to experimental improvements. To support this, the availability of a large number of Higgs bosons allows differential distributions to be measured, in particular at high $p_T$, which will also provide critical input for theory improvements. 

%\editnote{MD}{Heather/Elliot: See Andreas's comments about theory uncertainties. I don't think we should discuss this in great detail because this is needed for FCCee too. So I have just made a generic claim. See above. } %HG: I added a sentence specifically after the ratios, to call them out on this: provided that theoretical uncertainties can be sufficiently understood..

\begin{table}[h]
    \centering
    \begin{tabular}{lccccc}
    \hline\hline
         Kappa [\%] & HL-LHC & HL-LHC+FCC-ee & HL-LHC+FCC-hh \\
        \hline
        $\kappa_W$ & 1.6 & 0.38 & 0.39 \\
        $\kappa_Z$ & 1.6 & 0.14 & 0.63 \\
        $\kappa_g$ & 2.4 & 0.88 & 0.74 \\
        $\kappa_\gamma$ & 1.8 & 1.2 & 0.56 \\
        $\kappa_{Z\gamma}$ & 6.8 & 10. & 0.89 \\
        $\kappa_c$ & --  & 1.3 & -- \\
        $\kappa_t$ & 3.4 & 3.1 & 0.99 \\
        $\kappa_b$ & 3.6 & 0.59 & 0.99 \\
        $\kappa_\mu$ & 3.0 & 3.9 & 0.68 \\
        $\kappa_\tau$ & 1.9 & 0.61 & 0.9 \\
        \hline\hline
    \end{tabular}
    \caption{Higgs Kappa results for three scenarios: The HL-LHC, the HL-LHC plus FCC-ee only and the HL-LHC plus FCC-hh at 100 TeV only. Taken from Ref~\cite{HLLHCUpdate} and Tables 3 and 30 from Ref~\cite{de_Blas_2020}.}
    \label{tab:higgsKappas}
\end{table}

\subsection{Di-Higgs}

The physics motivations to better understand the Higgs self-coupling are numerous and profound. Investigating the self-coupling represents a unique opportunity to probe beyond the Standard Model (BSM) physics, as many theories predict deviations from the SM value. It is also crucial for investigating the electroweak phase transition, offering an ultimate test of the Higgs mechanism and its role in the stability of the electroweak vacuum. The precise value of the self-coupling directly determines whether our universe exists in a metastable state, connecting collider physics to fundamental questions of vacuum stability. Ultimately, a deep understanding of the self-coupling can potentially shed light on key unresolved questions such as the explanation of baryon asymmetry in the universe, as only a first-order electroweak phase transition could generate the conditions needed for electroweak baryogenesis. These capabilities make it an invaluable tool for advancing our understanding of fundamental physics.

\begin{figure}
    \centering
    \includegraphics[width=0.48\textwidth]{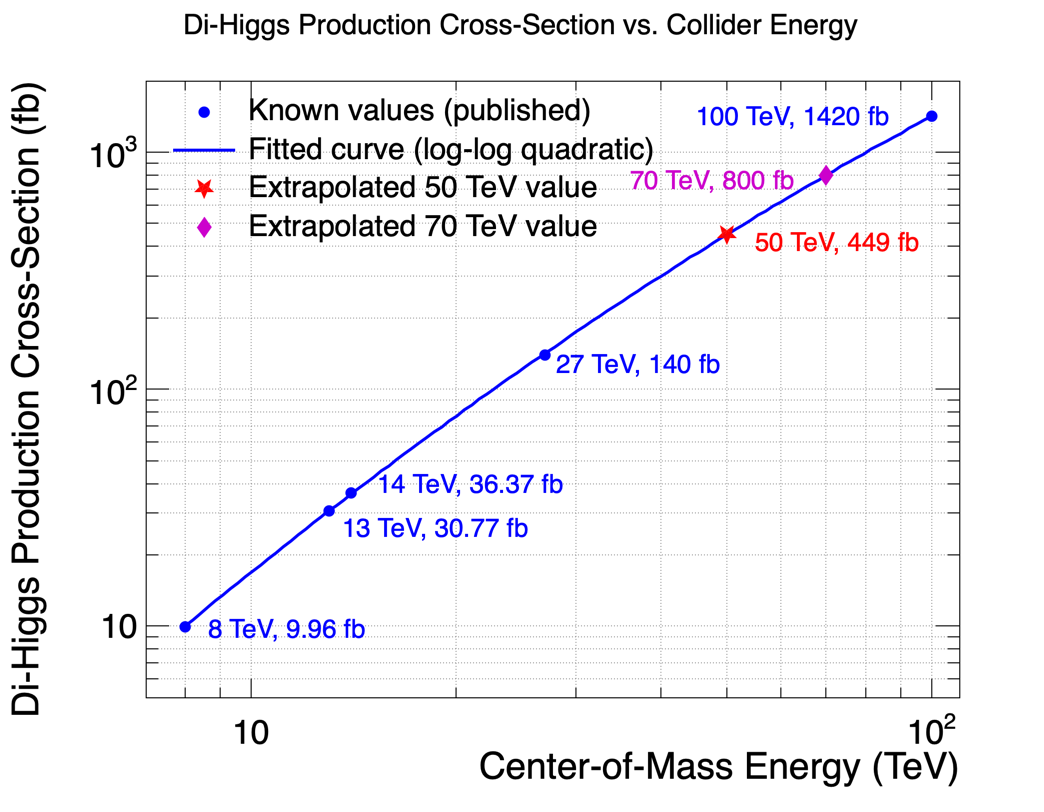}
     \includegraphics[width=0.48\textwidth]{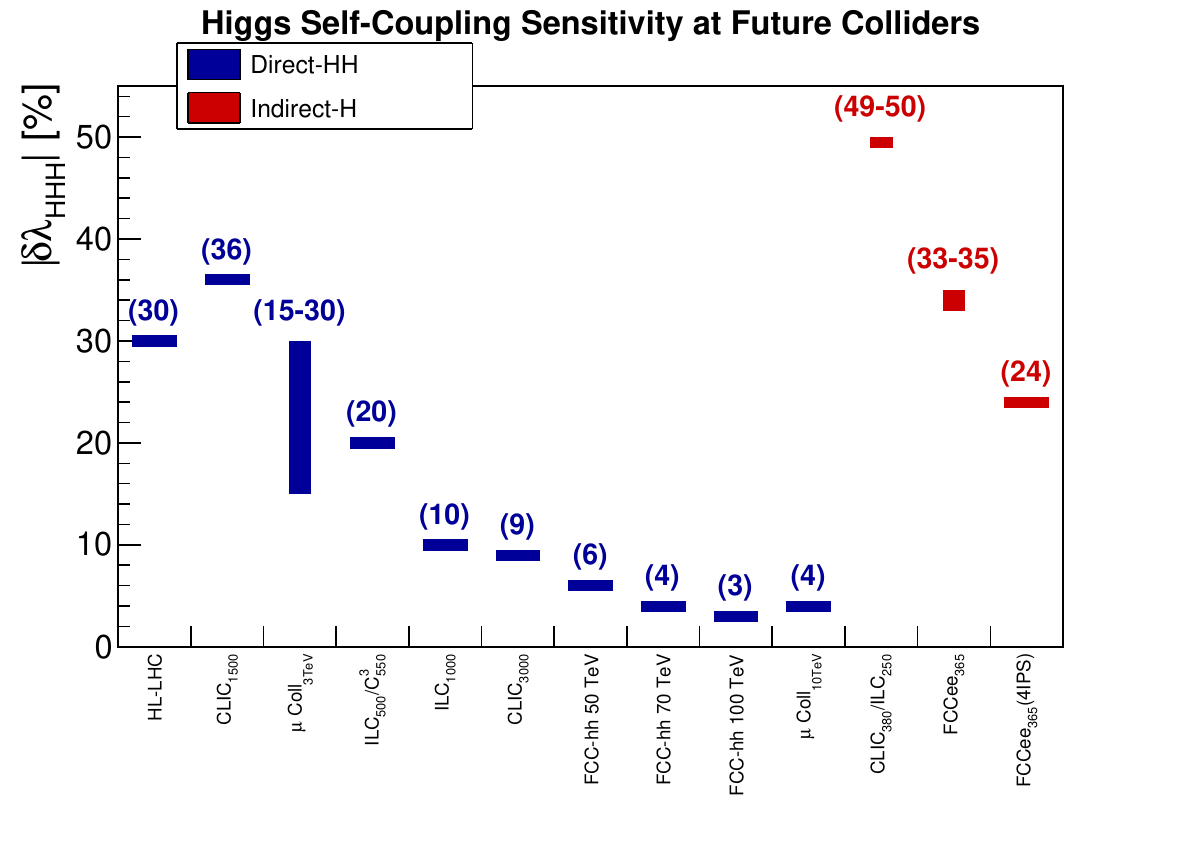}
    \caption{Left: Di-Higgs production cross-section as a function of centre-of-mass energy. The values of 50 and 70 TeV are extrapolated using a quadratic fit. The blue points are from Ref.~\cite{LHC_Higgs_WG4_YR4}. Right: Expected Higgs self-coupling sensitivity at future colliders, based on Ref.~\cite{Muhlleitner_DESYFutureCollPhys_2024} with updated numbers. The FCC-hh 50 and 70 TeV points are extrapolated.}
    \label{fig:dihiggs}
\end{figure}

A 70 TeV (50 TeV) hadron machine would enable measurement of the Higgs self-coupling to a precision of approximately 4\%  (6\%), compared to 3\% at the full 100 TeV FCC-hh\cite{Mangano:2020sao} and 30\% at the HL-LHC. The 50 TeV and 70 TeV results are extrapolated from the 100 TeV result using the square root of the expected number of di-Higgs produced. An extrapolation from the HL-LHC projections would give a smaller predicted uncertainty.  Such results would transform our understanding of the structure of the Higgs potential. The cross-section for Higgs-pair production would be approximately 12 times larger at 50 TeV and 25 times larger at 70 TeV than at the HL-LHC, as shown in Fig.~\ref{fig:dihiggs}, allowing for detailed studies of various di-Higgs production and decay channels. This statistical advantage enables precision measurements across multiple channels.

\subsection{Absolute Normalization}

Despite the production uncertainties related to QCD calculation and parton distribution functions (PDFs), it is possible to determine absolute normalisations in hadron colliders. Reference \cite{lipeles_2025_z6fgw-w5n60} shows that the ratio of Vector Boson Fusion (VBF) Higgs to $WW^* \rightarrow e\nu\mu\nu$ over non-resonant VBF $qq \rightarrow qqWW^* \rightarrow e\nu\mu\nu$ can be measured with low background at the 1-2\% level. In the context of the kappa framework, the numerator of the ratio is proportional to $\frac{\kappa_W^4}{\kappa_H^2}$, while the denominator is related to already well-measured electroweak couplings. There is a second-order effect in the denominator related to off-shell and $t$-channel Higgs exchange, but that introduces an opportunity to determine the absolute Higgs width itself \cite{Campbell:2013wga}. 

The theoretical production uncertainties are very similar between the numerator and denominator processes so they would not be expected to be a major limitation on the ratio. Similarly, the detector signals primarily differ in angles and modest shifts in energy distribution, so detector-related uncertainties should be similarly small. This leaves the background modelling where there is a plethora of control regions available. A 1-2\% uncertainty on $\frac{\kappa_W^4}{\kappa_H^2}$ would set a constraint on overall $\kappa$ scale of 0.5-1\%, which is similar to the scale of the $e^+e^-$ $\kappa$ constraints in Table \ref{tab:higgsKappas}. Additional, complementary constraints on absolute normalization can also be obtained from similar ratios such as $WH \rightarrow l\nu bb$ over  $WZ \rightarrow l\nu bb$ which is estimated to give 0.5\% sensitivity range\cite{Mangano:2681378}.

\section{Searches for the Direct Production of New Physics}
\label{direct_production}
As the LHC has so far not found signs of physics beyond the SM, but has excluded large regions of parameter space where new particles may have been. While never guaranteed, a new hadron collider, in contrast to $e^+e^-$ machines, would open up a large new discovery potential for direct production. The final answer of how much better a 100~TeV hadron collider is than a 50~TeV one can only be answered by nature as it depends on the energy scale of BSM physics, but the relative energy reaches can be understood. The case for TeV-scale BSM physics still remains strong and being able to explore significant amounts of new phase-space as soon as possible should not be dismissed.

\begin{figure}
    \centering
    \includegraphics[width=0.52\textwidth]{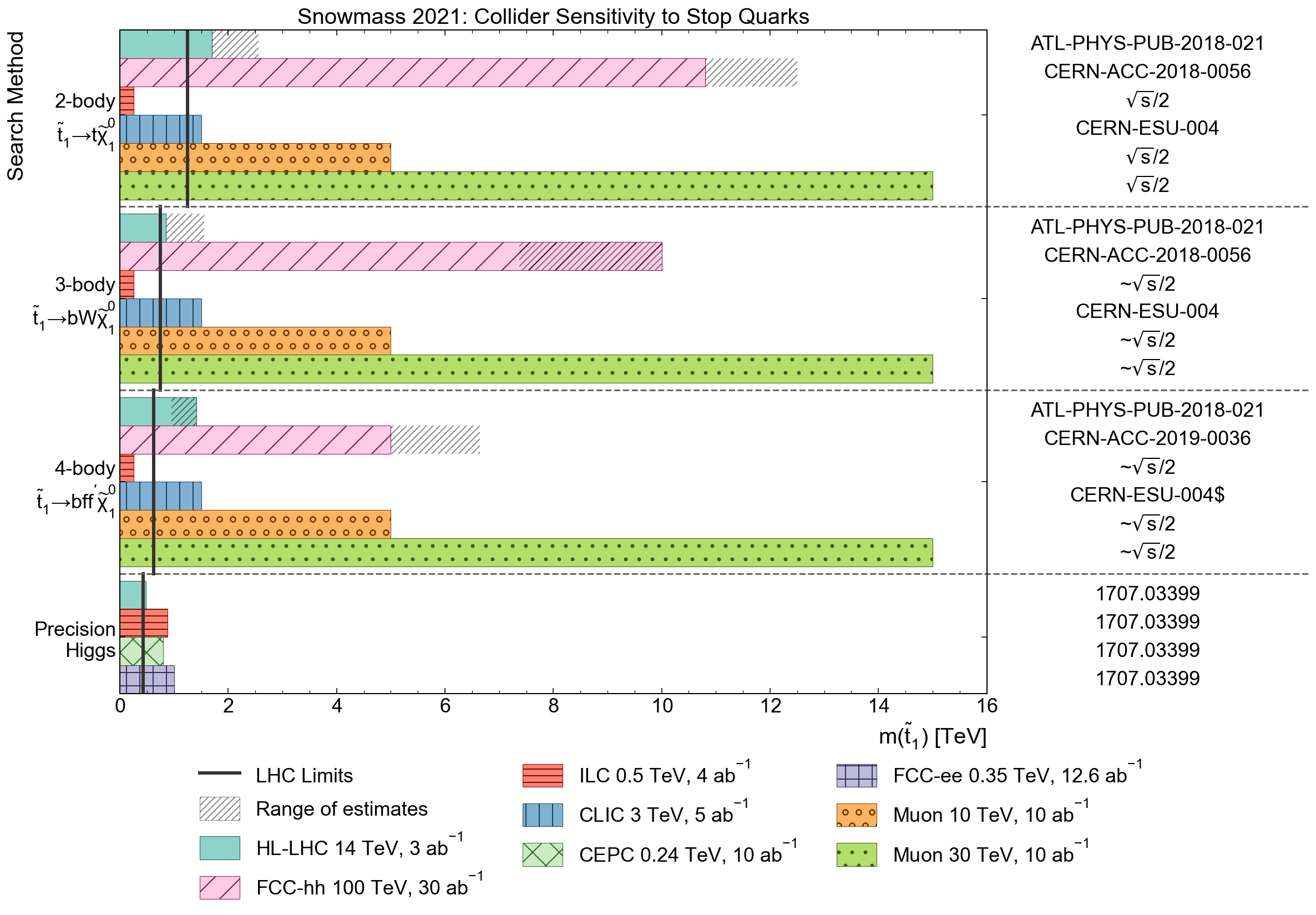}
     \includegraphics[width=0.41\textwidth]{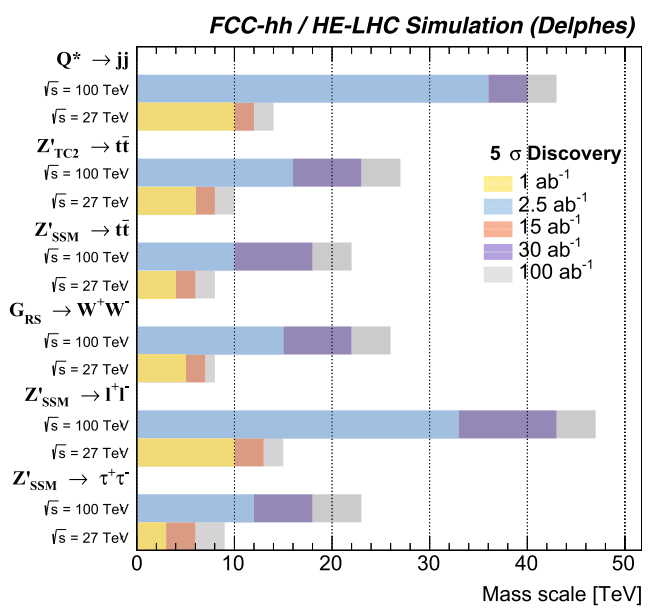}
    \caption{Left: Estimated stop exclusion reaches for various colliders and search methods, from Ref.~\cite{Bose:2022obr}. The limits are categorized based on the mass difference into two-body, three-body and four-body decays. The bars represent the maximum excluded stop mass ($m(\tilde{t}_1)$) in each region. Precision Higgs constraints are derived from deviations in Higgs production rates under the assumption that stops are the only source of BSM effects.
     Right: Summary of the $5\sigma$ discovery reach as a function of the resonance mass for different luminosity scenarios of FCC-hh and HE-LHC From Ref.~\cite{Helsens:2019brx} .}
    \label{fig:directnewphys}
\end{figure}

Figure \ref{fig:directnewphys} shows the very large gain in reach for new hadron colliders in two representative model classes,  new top partners (left) and new resonances (right). Stop-squark searches would move from HL-LHC limits around 2 TeV to 8-10 TeV for a new collider (depending on specific energy and luminosity), and SSM Z' prime searches would move from 5 TeV to 33-46 TeV. These examples show how the hadron collider opens up a new energy regime above the electroweak scale, even for the challenging compressed scenarios shown in Figure \ref{fig:directnewphys} (left).

\begin{figure}
    \centering
    \includegraphics[width=0.5\linewidth]{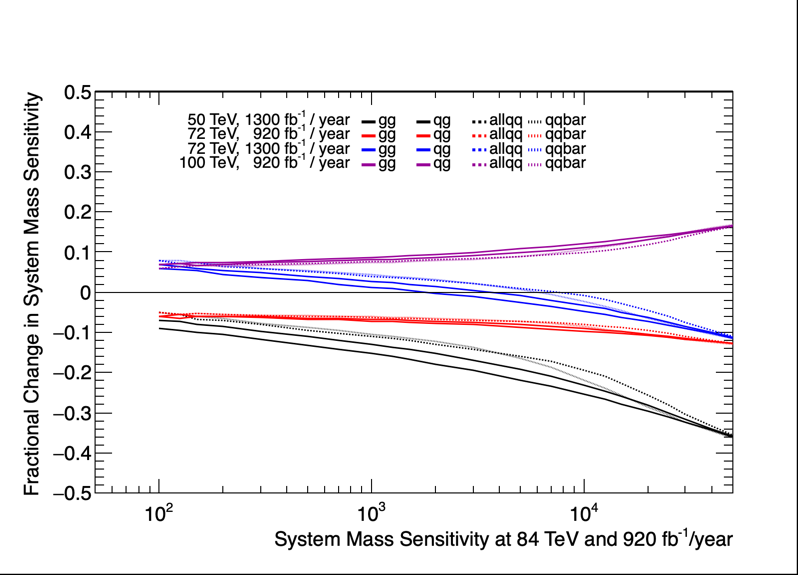}
    \caption{Comparison of system mass reach calculated with the Collider Reach Tool \cite{collidertool} (see text) for the collider scenarios shown in Table \ref{tab:Zimmermann}. For a given expected mass reach (either 95\% CL or $5\sigma$ discovery) with respect to a nominal 84 TeV 920\,ab$^{-1}$/year machine on the $x$-axis, the plot gives the fractional shift in expected mass reach compared to that nominal machine on the $y$-axis for other collider scenarios. This shift depends weakly on what type of partons initiate the signal process: gluon-gluon $gg$, quark-antiquark $q\bar{q}$, or quark-gluon $gq$, and flavour-independent quark-quark $qq$.} 
    \label{fig:energycomp}
\end{figure}

The rough BSM energy reach of collider scenarios can be compared, including the luminosity differences, using the {\sc Collider Reach} tool~\cite{collidertool}. This calculates where an equivalent number of signal events would be produced based on the parton luminosities, which is sufficient to understand the broad impacts of energy and luminosity. The results shown in Figure \ref{fig:energycomp} can then be used to estimate the gain/loss of sensitivity compared to the nominal 84 TeV, 920 fb$^{-1}$/year scenario. For example, comparing F12PU to F14 for system masses below 3\,TeV there would be a gain in lowering the energy and raising the luminosity, while above 3\,TeV there would be some losses in sensitivity. Specifically, if F14 would set limits at 40\,TeV (near the right edge of the plot) for a particular a $Z'$ , the F12PU would set a limit at 36\,TeV, i.e a 10\% loss in the energy reach. Therefore, depending on the system mass of interest, there are trade-offs between higher energy and more luminosity.
%\editnote{MD}{Elliot: Comment from Karl that this figure needs more description in caption and text... done}

%\editnote{AL}{Add something about DM candidate reach?}

\section{Non-resonant Production and Effective Field Theory Interpretations}
\label{eft}

Deviations from SM expectations in precision measurements could indicate BSM physics. Such deviations could impact multiple measurements, so understanding the precision on a single parameter alone is not sufficient. The precision on various parameters must be compared in a consistent theoretical framework. If the new physics energy scale is significantly higher than the precision measurements, an effective field theory (EFT) with higher dimension operators ($\geq$5) can be used to represent a complete set of possible deviations from the Standard Model. One such EFT is the SMEFT, which makes some theoretical assumptions about the nature of the new physics, but gives a broad context for the comparisons of measurements~\cite{Brivio:2017vri}. One key feature of such EFTs is that the BSM effects grow generally with energy squared~\cite{Farina:2016rws}. Qualitatively that means a 0.1\% measurement at 100~GeV is roughly equivalent to a 10\% measurement at 1~TeV.

Recently, progress has been made on such comparisons~\cite{Celada:2024mcf}, but the inputs to such analyses are limited by the level of analysis work that has been conducted both for the current LHC experiments and in the context of the FCC-hh proposal. This was noted in Ref.~\cite{Belloni:2022due}: `At this point, not enough information was available to include pp colliders beyond the LHC (such as HE-LHC or a O(100)-TeV collider) in the global fit. It is likely that these machines have superior sensitivity to many energy-dependent operators, such as 4-fermion operators involving quarks and several operators that mediate multi-boson interactions'~ The LHC results and expectations presented in the previously mentioned summaries are necessarily only those which were complete at the time of compilation. This has led to a significant underestimate of the power of high-energy precision Standard Model measurements at hadron colliders. 

\begin{figure}
    \centering
    \includegraphics[width=0.5\linewidth]{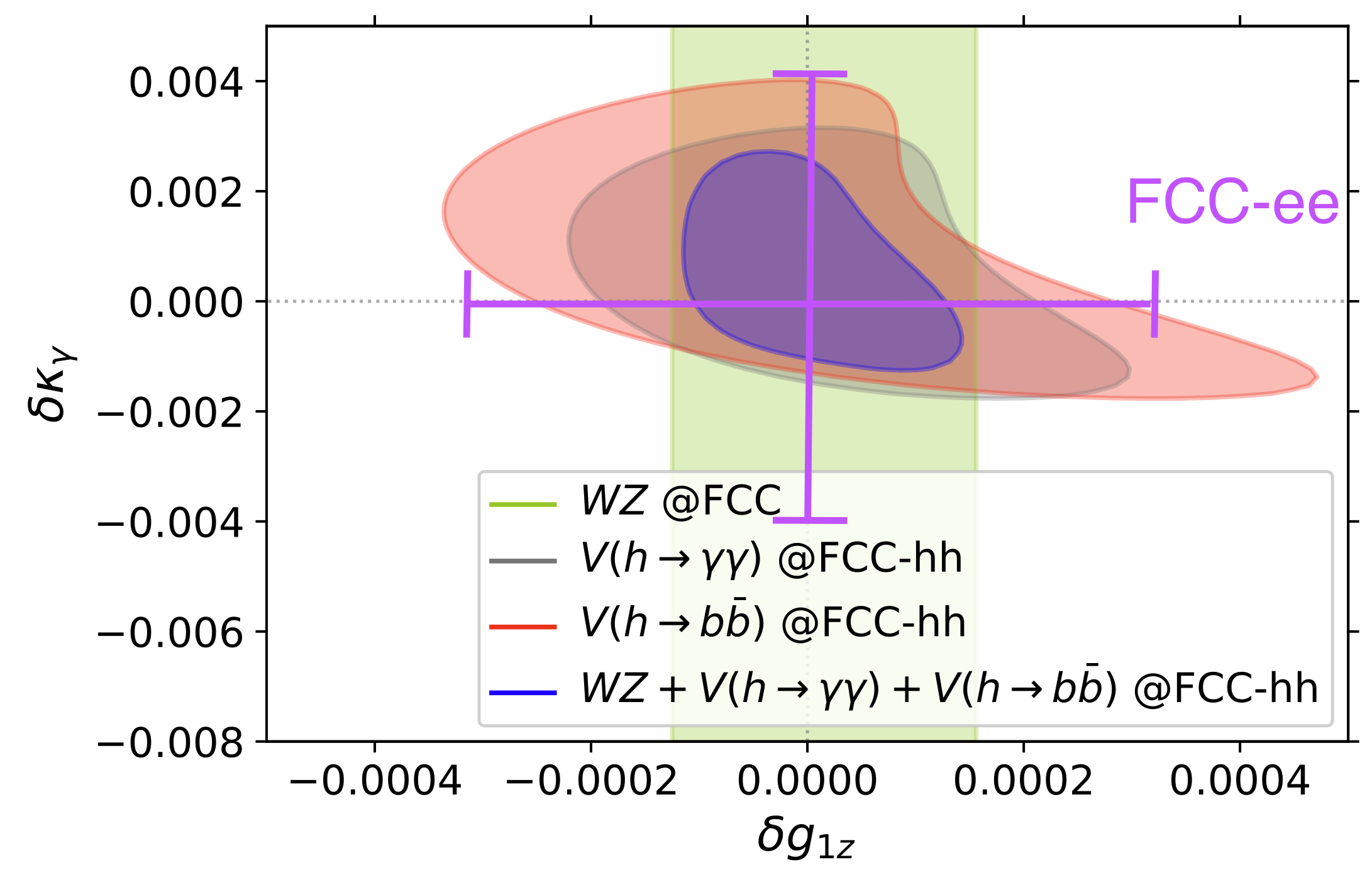}
    \caption{Comparison of sensitivities for $\delta k_\gamma$ vs $\delta g_{1z}$ from FCC-hh at 100 TeV \cite{Bishara:2022vsc} with  the FCC-ee sensitivity\cite{FCC:2018byv} overlaid in purple.}
    \label{fig:aTGC}
\end{figure}

An example of this is the state and progress of measurements of triple gauge couplings. The modifications of the triple-gauge couplings can be parametrised in a variety of ways. Here, we present a comparison of the one-parameter $\delta g_{1z}$ selected based on the current availability of measurements and predicted sensitivities. The estimates collected by Snowmass gave ultimate HL-LHC sensitivity as $\sim 6\cdot 10^{-2}$ and $\sim 3\cdot 10^{-4}$ for the FCC-ee program \cite{Belloni:2022due}. At the time that report was written, CMS had already reported a sensitivity of $\sim 7\cdot 10^{-2}$ with just 35~fb$^{-1}$, approximately a 100th of the ultimate HL-LHC dataset, and an early analysis technique. The source of the discrepancy is primarily the choice of final state. The Snowmass report (as well as the FCC-ee CDR\cite{FCC:2018byv}) based the HL-LHC prediction of an analysis of the fully-leptonic decays of $WZ$ production, while the CMS result uses semileptonic decays of $WZ$ production. Estimates of the sensitivity for some other relevant final states have been made for FCC-hh\cite{Bishara:2022vsc} at 100\,TeV and 30\,ab$^{-1}$. Figure \ref{fig:aTGC} compares these to estimates from the fits to the full FCC-ee program\cite{FCC:2018byv}. The results show that the effective precision of the hadron collider is comparable if not better. A detailed and more comprehensive study is required to understand how the $e^+e^-$ programme sensitivities compare to hadron collider sensitivities, both for HL-LHC, which may provide much stronger sensitivity than currently estimated and for a future hadron collider, which may cover or exceed the  $e^+e^-$ proposals, i.e. a clear case for the uniqueness of the $e^+e^-$ machine needs to be made.

\section{Precision Versus Direct Production} 
\label{precisionvproduction}

\begin{figure}[h]
    \centering
    \begin{subfigure}[t]{0.48\textwidth}
         \centering
         \includegraphics[width=\linewidth]{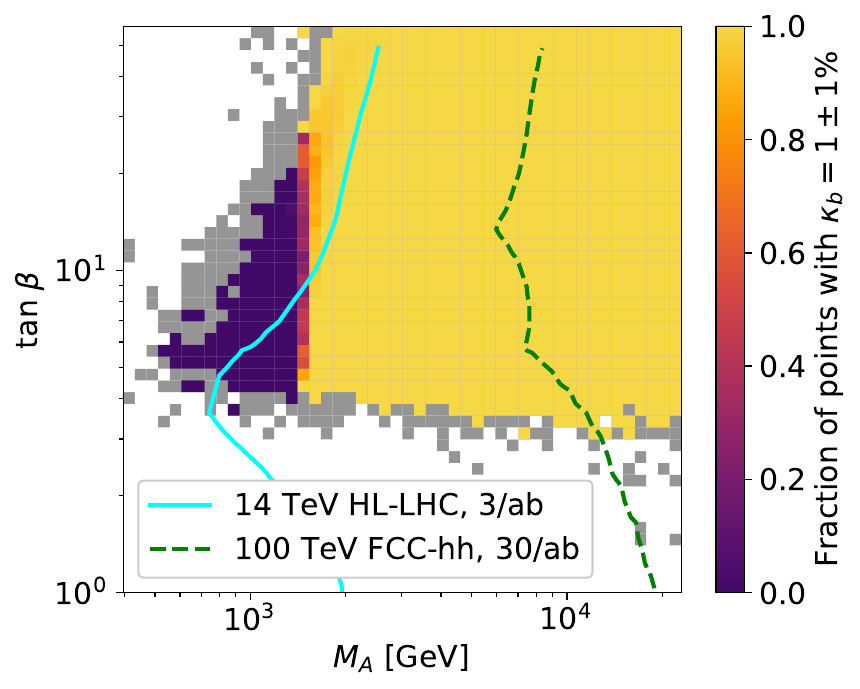}
    \caption{Fraction of models in a scan over SUSY parameter space in the RPC pMSSM that have a $\kappa_b$ deviation from the SM exceeding 1\%. The dark area would be excluded roughly 95\% by the $e^+e^-$ precision measurements. The light blue and dotted green lines show expected exclusions from HL-LHC and FCC-hh at 100 TeV respectively.  }
    \label{fig:2hdm}
    \end{subfigure}
    \hfill
    \begin{subfigure}[t]{0.48\textwidth}
         \centering
         \includegraphics[width=\linewidth]{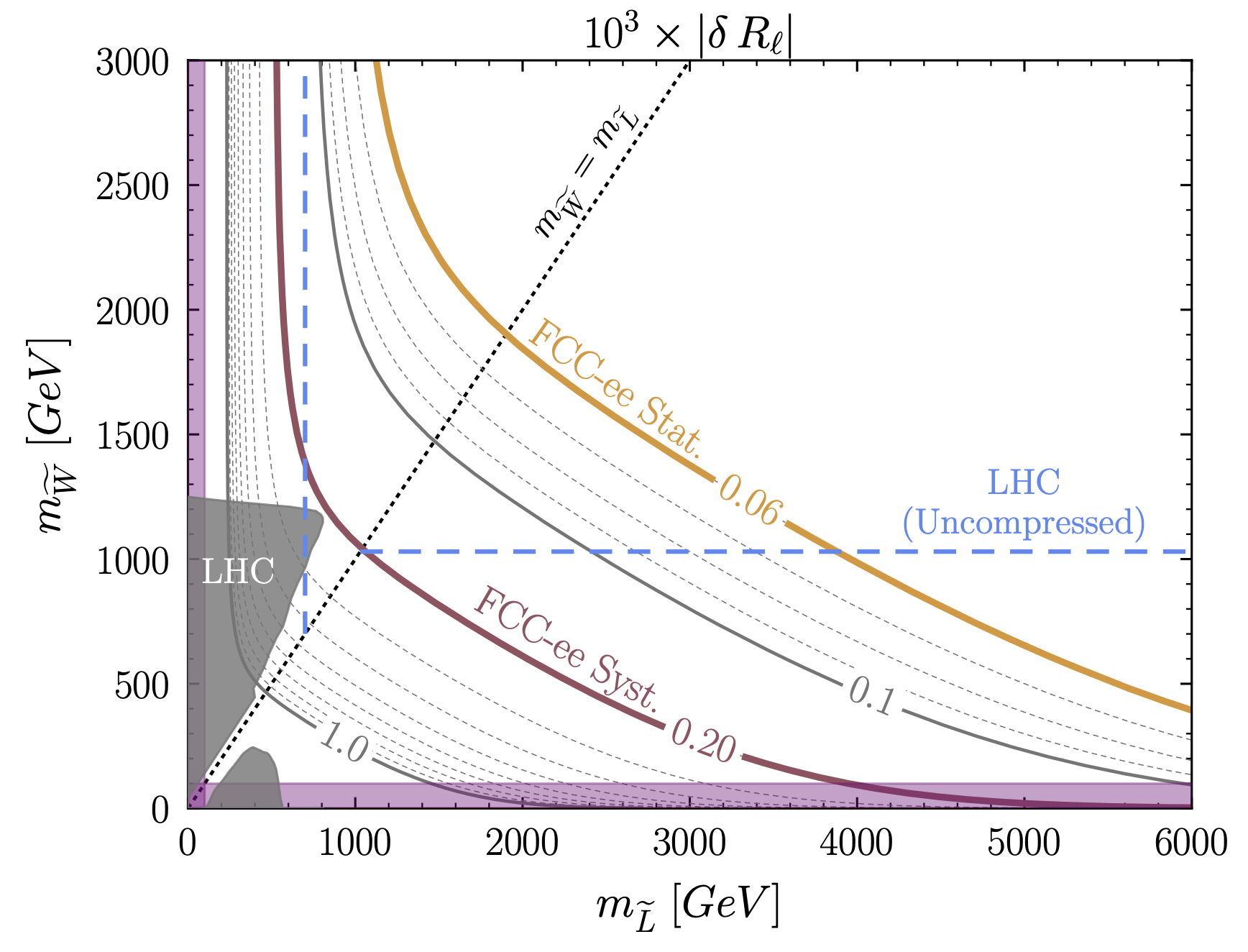}
    \caption{Comparisons of constraints on sleptons from an FCC-ee prediction for the measurement of $R_\ell = \frac{\Gamma(Z\rightarrow\mathrm{hadrons})}{\Gamma(Z\rightarrow\ell\ell)}$ and current LHC limits~\cite{Knapen:2024bxw}. The dotted blue lines and grey shaded areas show existing LHC limits.}
    \label{fig:Rlexample}
    \end{subfigure}
    \caption{Example comparisons of precision constraints with direct searches at hadron colliders.}
\end{figure}

Precision measurement constraints on new physics depend strongly on the model being considered.
In the context of EFTs, the constraints scale as $\frac{c}{\Lambda^2}$ where $c$ is coupling constant and $\Lambda$ is the energy/mass scale of the new physics. If $c$ is large, this scale can be constrained beyond the direct production reach of hadron colliders (although they may still have better constraints on the EFT operators). If $c$ is not large, then the constraints are relatively weak.  The bottom set of bars in Figure \ref{fig:directnewphys}(left) shows limits calculated from $h\rightarrow\gamma\gamma$ and $h\rightarrow gg$ loops on top partners~\cite{Essig:2017zwe}. The future collider projects are not competitive with HL-LHC. Figure \ref{fig:2hdm} shows the fraction of models in a scan over SUSY parameter space in the RPC pMSSM that have a $\kappa_b$ deviation from the SM exceeding 1\%. Only a small part of that region will not be covered by the HL-LHC, and a future hadron collider would far exceed the
precision measurement's sensitivity. Finally, Figure \ref{fig:Rlexample} shows a recent study of constraints from a Tera-Z program measurement of $R_\ell =\frac{\Gamma(Z\rightarrow\mathrm{hadrons})}{\Gamma(Z\rightarrow\ell\ell)}$ \cite{Knapen:2024bxw}. Even in the context of SUSY, there is a lot of model dependence. The authors find in two of three scenarios considered, the LHC or expected HL-LHC limits exceed the precision sensitivity with the except of some compressed regions of parameter space, and the third model is an $R$-parity violating scenario which does not have LHC limits for first and second generation couplings.

%\editnote{MD}{Elliot: Karl suggested more description for the left plot. I am also confused by what is meant by 'dark area' because the color code is the fraction. It is this sentencce that confuses me 'The dark area would be excluded roughly 95\% by the $e^+e^-$ precision measurements' can we rephrase it? }

\section{Impact on the Community} 
\label{community}

We strongly support the idea of a flagship collider at CERN as the centre of the experimental particle physics community. The physics potential of current and future projects is the foundation of our community. 
A dynamic programme is required to attract and retain young outstanding talent in the field, and to maintain the relevant expertise for designing, building, and operating accelerators and detectors. The timescale of an FCC-hh high-energy machine after the retirement age of our current graduate students will have (and potentially already has) a significant impact on their motivation and interest in the field. As the physics scope of an FCC-hh machine is significantly broader than $e^+e^-$ machines and able to accommodate a wider range of physics interests, if we are able to put forward a project on a much shorter timescale (e.g. 2045-2060) this would re-energize the collider community, in particular those interested in direct searches for BSM physics. An near-term hadron machine at CERN would also support the LHC expertise in the CERN accelerator division, which would otherwise be hard to maintain. An near-term machine will impact the whole community from accelerator physicists, experimental particle physicists and also particle theorists, representing thousands of individuals, and hundreds of institutions.

\section{Conclusion}
%\noindent \lipsum[1] \cite{1}
A near-term intermediate-energy hadron collider would provide a broad and comprehensive program for exploring physics at the electroweak scale and providing a window beyond without needing an $e^+e^-$ machine. Specifically, it simultaneously provides a powerful Higgs and electroweak precision program, as well as a direct (and EFT) probe program.

Bringing this opportunity into the career span of the current younger generation of scientists is of critical importance to
maintain the vitality of the field, as well as the required expertise. 

%To achieve this goal, the FCC program should layout a decision-making process that evaluates in detail options for proceeding directly to a hadron collider including
%possibilities of reducing energy targets and staging the magnet
%installation to spread out costs.

%Alternate last sentence:
To achieve this goal, the FCC program should lay out a decision-making process that evaluates in detail options for proceeding directly to a hadron collider, including the possibility of reducing energy targets and staging the magnet installation to spread out the cost profile.

%Sentence from Viviana could be worked in.
%While any future collider represents an advancement, the steep scaling of physics reach with energy makes a compelling case for pursuing the highest practical energy, with each increment providing substantial returns on scientific investment for understanding electroweak symmetry breaking and its cosmological implications.

% To achieve this goal, a staging program including two hadron colliders
% should be studied in detail with a clear decision process, including
% more comprehensive physics performance comparisons, possibilities of
% reducing energy targets, and staging the magnet installation to spread
% out costs.

\bibliographystyle{unsrt}
\bibliography{references}
% HG: Use bibtex
%\begin{thebibliography}{99}

%\bibitem{1} Spiegel, M. R. (1981). Theory and problems of Advanced Calculus: Si (metric) edition. McGraw-Hill. 

%\end{thebibliography}

\appendix
\renewcommand{\thesection}{Appendix~\Alph{section}}
\section*{Appendix: US community interest and outlook on the near-term hadron collider as an alternative scenario }

Editors: Viviana Cavaliere(BNL), Heather Gray(UC Berkeley and LBNL), Elliot Lipeles (University of Pennsylvania)\\

\noindent Anthony Affolder\textsuperscript{1},
Haider Abidi\textsuperscript{2},
John Alison\textsuperscript{3},
Prachi Arvind Atmasiddha\textsuperscript{4},
Michael Begel\textsuperscript{2},
Sully Billingsley\textsuperscript{5},
Elizabeth Brost\textsuperscript{2},
Jared Burleson\textsuperscript{6},
Yuan-Tang Chou\textsuperscript{7},
Angelo Di Canto\textsuperscript{2},
Thiago Costa de Paiva\textsuperscript{8},
Emily Duden\textsuperscript{9},
Ian Dyckes\textsuperscript{10},
JiJi Fan\textsuperscript{11},
Mike Hance\textsuperscript{1},
Philip Harris\textsuperscript{12},
Timon Heim\textsuperscript{10},
Christian Herwig\textsuperscript{13},
Calla Hinderks\textsuperscript{14},
George Iakovidis\textsuperscript{2},
I.~Joseph Kroll\textsuperscript{4},
Matt LeBlanc\textsuperscript{11},
Verena Ingel Martinez Outschoorn\textsuperscript{8},
Christopher Meyer\textsuperscript{15},
Mark Neubauer\textsuperscript{6},
Jason Nielsen\textsuperscript{1},
Katherine Pachal\textsuperscript{16},
Dylan Rankin\textsuperscript{4},
Jennifer Roloff\textsuperscript{11},
Benjamin John Rosser\textsuperscript{17},
Andrea Sciandra\textsuperscript{2},
Jeffrey Shahinian\textsuperscript{4},
Hayden Shaddix\textsuperscript{18},
Punit Sharma\textsuperscript{2},
Louise Skinnari\textsuperscript{19},
Scott Snyder\textsuperscript{2},
Giordon Stark\textsuperscript{1},
Ryszard Stroynowski\textsuperscript{5},
Kaito Sugizaki\textsuperscript{4},
Max Swiatlowski\textsuperscript{20},
Emily Thompson\textsuperscript{10},
Haichen Wang\textsuperscript{20},
Chenwei Zhao\textsuperscript{20}\\

\noindent \textsuperscript{1}University of California, Santa Cruz (SCIPP), 
\textsuperscript{2}Brookhaven National Laboratory (BNL), 
\textsuperscript{3}Carnegie Mellon University, 
\textsuperscript{4}University of Pennsylvania, 
\textsuperscript{5}Southern Methodist University, 
\textsuperscript{6}University of Illinois at Urbana–Champaign, 
\textsuperscript{7}University of Washington, 
\textsuperscript{8}University of Massachusetts Amherst, 
\textsuperscript{9}Brandeis University, 
\textsuperscript{10}Lawrence Berkeley National Laboratory (LBNL), 
\textsuperscript{11}Brown University, 
\textsuperscript{12}MIT, 
\textsuperscript{13}University of Michigan, 
\textsuperscript{14}University of Oregon, 
\textsuperscript{15}Indiana University, 
\textsuperscript{16}TRIUMF, 
\textsuperscript{17}University of Chicago, 
\textsuperscript{18}Northern Illinois University, 
\textsuperscript{19}Northeastern University, 
\textsuperscript{20}UC Berkeley (and LBNL)

\subsection*{Near-term hadron collider }

Because of the wide range of interests among various experts in the U.S. community and the lack of a formal mechanism for prioritization no prioritization of the list of alternatives was given in the U.S. inputs to the European Strategy Group. This document is a statement by a subset of the U.S. community about the need to prioritize R\&D related to a near-term cost-optimized hadron collider as a potential alternative plan should the FCC-ee prove to be infeasible or not competitive.  Scenarios where a change of direction may be required include competition from other projects, technical performance considerations of the preferred option, the HL-LHC outperforming expectations, and potential discoveries at the HL-LHC or other experiments that point to an energy scale and class of models where a different approach may be needed. A list of the endorsements collected is at the end of the document.

Cost optimization could include a variety of changes to the design and scope. For example, a reduction in center-of-mass energy to 70 TeV would save substantial cost and reduce research and development time, while having only a limited (and not entirely negative) impact on physics performance compared to the current FCC-hh baseline energy of 84 TeV. Such an alternative plan aligns with the preferred option, since the R\&D required is essentially the same as for the full FCC integrated program. It also builds on the significant investments made by CERN, the member and associate member states, and other international partners over the past years in site-specific feasibility studies that laid the groundwork for developing a 91 km circular tunnel in the French-Swiss region. Moreover, to preserve this alternate plan, a focus on magnet technology development and lattice configuration alternatives is needed. Unlike other options under consideration, this pathway advances the core technological bottleneck for future energy frontier machines, supports timely civil construction, preserves programmatic momentum, and offers unmatched physics reach,  potentially achieving physics operations in the early 2050s. It is also synergistic with the ongoing magnet R\&D development program in the United States and Europe and allows for an efficient use of existing resources and efforts if a feasibility study is undertaken. 

\subsection*{Physics Case}

A 70 TeV hadron collider [https://arxiv.org/abs/2504.00951] would produce approximately half a billion Higgs bosons annually—making it the ultimate Higgs factory. While they do not achieve the same level of the same precision in model-independent Higgs coupling measurements of electron-positron machines, the hadron collider achieves sub-percent precision on major Higgs couplings through ratio measurements, provides unique access to rare production and decay modes, and enables differential measurements at high transverse momentum where new physics effects are enhanced. The flagship measurement, the Higgs self-coupling, can be measured to 4-8\% precision at 70 TeV (compared to 27-30\% at HL-LHC and FCC-ee), providing crucial tests of electroweak symmetry breaking and vacuum stability.

High-energy hadron colliders have far superior sensitivity to the direct production of new physics than any lower-energy precision machines. This provides discovery reach for Z' bosons to 18-25 TeV mass, supersymmetric particles and vector-like quarks to 3.5-7 TeV mass ranges, and other exotic particles. The direct discovery reach of a 70 TeV collider is reduced, typically by 10-20\% compared to an 85 TeV machine. Among all proposed alternate plan options, only a hadron collider maintains substantial discovery potential for directly produced new physics beyond the HL-LHC reach. Such breadth of scientific discovery capability and characterization ability any new physics discovered cannot be replicated by precision-only programs. 

Precision measurements of Standard Model processes at multi-TeV energies provide powerful constraints on new physics through effective field theory frameworks. Since EFT sensitivity scales with the square of the energy, a 1 TeV measurement at a hadron collider can match or exceed the indirect sensitivity of much more precise measurements at lower energies. Recent studies suggest that high-energy diboson production, vector boson scattering, and top quark processes at a 50-70 TeV collider may achieve EFT reach competitive with or exceeding electron-positron collider programs for many operator classes, while maintaining the unique capability for direct discovery if new physics appears at accessible mass scales.

\subsection*{Strategic Alignment and Synergies}

The feasibility study for a near-term hadron collider focuses precisely on the magnet technology development—Nb$_3$Sn and High Temperature Superconducting (HTS) systems—that forms the critical path for the ultimate FCC-hh vision. This is not ancillary work but rather the precise R\&D program already identified as essential for advancing CERN's long-term strategic roadmap. Investing in this direction, therefore, serves dual purposes without distracting from the on-going FCC-ee development efforts. 

Moreover, it allows the creation of an alternative path should another country decide to build an electron-positron machine. A near-term hadron collider would allow CERN to maintain undisputed global leadership, regardless of the developments in the competing electron-positron programs, maintaining access to the full spectrum of international collaborators and their expertise.

\subsection*{Programmatic and Financial Advantages}

This alternate plan is fully compatible with immediate civil construction initiation. The 91 km FCC tunnel and associated infrastructure can proceed on the current timeline regardless of whether the installed machine is FCC-ee or a FCC-hh. This maintains critical programmatic momentum, demonstrates tangible progress to funding agencies and political stakeholders, and most importantly, preserves the institutional commitment to the "FCC" program as CERN's next flagship initiative. This approach also provides natural decision points. As tunnel excavation progresses, the magnet technology development program will continue to mature, providing increasingly refined cost and performance projections that can inform the final selection.

The hadron collider pathway would maintain eligibility for the full portfolio of funding mechanisms cultivated for the FCC vision—EU structural funds, national government commitments, and private philanthropic contributions attracted by energy frontier science. Other proposed alternate options may struggle to generate equivalent enthusiasm among stakeholders who have invested in the vision of unprecedented energy reach. The continuity of programmatic identity ("FCC" branded infrastructure and science) substantially increases the likelihood of sustained political support across multiple budget cycles and governmental transitions.

\subsection*{Timeline and Cost }

A near-term hadron collider targeting the early 2050s for physics operations creates a significant gap between HL-LHC conclusion (mid-2040s) and the next-collider startup. However, this timeline offers strategic advantages for cost management through natural CERN budget accumulation during the interim period, allows magnet development to continue and its technology to reach full maturity, and provides adequate time for comprehensive civil construction without pressure on schedule that historically drives cost overruns. In addition, the analysis of the HL-LHC dataset is expected to extend for several years beyond the end of data-taking, ensuring workforce development of early career scientists to continue physics studies while simultaneously working on the technical challenges needed for the FCC program.
As the projected cost of an eventual FCC-hh is substantial, ongoing R\&D is heavily focused on cost optimization—particularly in magnet design, cryogenics, and tunnel infrastructure—with the aim of significantly reducing costs while maintaining physics performance. As noted in the preliminary conclusions on the assessment of the large-scale accelerator projects from the ESG “targeting a slightly reduced centre-of-mass energy with ~12 T dipoles, represents a possibly lower-cost pathway”. 

\subsection*{Conclusion}

A near-term hadron collider represents the optimal alternative plan to FCC-ee that aligns U.S. and European strategic interests. It fulfills the U.S.2023 P5 strategy and the long-term vision under the 2026 U.S. National Academies of Sciences’ Elementary Particle Physics Report for timely energy frontier physics while respecting the realities of magnet technology maturation, addresses competitive positioning relative to international electron-positron projects, maintains civil construction momentum, and delivers the strongest physics case among feasible alternatives considered at CERN.

\end{document}